\documentclass[prl, preprint]{revtex4-1}
 
\usepackage{graphics}
\usepackage{graphicx}
\usepackage{pstricks}
\usepackage{amsbsy}
\usepackage{amsmath}

\renewcommand{\vec}{\mathbf}

\renewcommand{\d}{\mathrm{d}}

\begin{document}

\title{A corrected single element Maxwell visco-elastic model}

\author{J.S. Hansen}
\address{"Glass and Time", IMFUFA, \\ 
Dept. of Science and Environment\\
Roskilde University, DK-4000 Denmark}

\begin{abstract}
	The original Maxwell visco-elastic constitutive model cannot  
	predict the correct mechanical properties for most fluids. In this work, 
	the model is generalized with respect to wave-vector and extended with a correction function.	
	This new model has only two free parameters and avoids the attenuation-frequency locking 
	present in the original model. Through molecular simulations it is shown that 
	the model satisfactory predicts the transverse dynamics of the 
	binary Lennard-Jones system at different temperatures,
	as well as water and toluene at ambient conditions.  From the correction function it is shown 
	that the viscous response is significantly reduced compared to the 
	predictions from the original model when taking the wave-vector dependency into
	account. Morover, a temperature dependent 
	characteristic length scale of maximum reduced viscous response is
	identified.
\end{abstract}	

\maketitle

It is fascinating how well continuum hydrodynamics can describe the dynamics of 
fluid systems on small length scale \cite{koplik:pfa:1989,travis:pre:1997,hansen:book:2022}. 
The limitations of the hydrodynamic theory  
depend on the specific dynamical properties, the fluid system one studies, 
as well as the set of constitutive models applied 
\cite{thien:book:2002,eijkel:mnf:2005,bryk:jcp:2010,bocquet:csr:2010,bocquet:lapchip:2014,hansen:molsim:2021}. 
An example of the latter is that Newton's law of viscosity for the fluid stress cannot 
predict the visco-elastic behavior observed for even for simple fluid systems on sufficiently small length scales 
\cite{alley:pra:1983,palmer:pre:1994}. A natural extension to the hydrodynamic description is then to 
apply generalized hydrodynamics (possibly in the Zwanzig-Mori frame work),
where the viscosity is wave-vector and frequency dependent 
\cite{levesque:pra:1973,hansen:book:2006}. Alternatively, 
one can use a visco-elastic constitutive model for the stress, 
for example, Maxwell's model \cite{maxwell:1867} which is the
focus here.  

The Maxwell model states that the total strain rate, $\dot{\gamma}$, is the sum of the 
viscous strain rate $\sigma/\eta_0$ and elastic strain rate $\dot{\sigma}/G$, where $\sigma$ is the 
stress, $\eta_0$ the shear viscosity, and $G$ the shear modulus (or modulus of rigidity). The 
standard dot-notation is here used to denote the time derivative. 
Maxwell's model can be expressed as
\begin{eqnarray}
	\eta_0 \dot{\gamma} = \left(1 + \tau_M  \partial_t \right)\sigma \, ,
\end{eqnarray}
where $\tau_M = \eta_0/G$ defines the Maxwell relaxation time. The simple linear combination between a purely 
viscous response (dissipation) and purely elastic response (energy storage),
will here be termed ideal mixing of viscosity and elasticity (or dissipation and storage). 

In a mechanical network representation, the Maxwell element is a 
dash pot (with viscosity $\eta_0$) serially connected to a spring (with characteristic time $\tau_M$). 
The Maxwell model itself can of course also be extended in different ways, for example, one can introduce a set 
of Maxwell elements connected in parallel (a Maxwell-Weichert element) \cite{tschoegl:book:1989}; 
each element is then characterized by a viscosity and characteristic time. 
This leads to a large model parameter space naturally allowing for better agreement with experimental
data. Moreover, the Maxwell model can be generalized by allowing both $\eta_0$ and $\tau_M$ 
to be functions of wave-vector, that is, the viscous and elastic responses become length scale dependent 
\cite{hansen:book:2006}.  
Mizuno and Yamamoto \cite{mizuno:prl:2013} introduced a combination of two parallel Maxwell elements 
with wave-vector dependent viscosities and Maxwell times. They showed that introducing the 
concepts of slow and fast relaxations their four parameter model featured 
good agreement with molecular dynamics data for a viscous model system. 

In this Letter, it is shown that the generalized, i.e., wave-vector dependent, 
single element Maxwell model fails to predict the correct transverse 
hydrodynamics for even simple systems in the visco-elastic regime. 
The deviation from simulation data is quantified by simply relaxing the ideal mixing rule 
through a correction function. The physical interpretation of this function 
is that the fluid viscous property is not determined by the zero-frequency viscosity, 
but also by other mechanical storage processes than what the Maxwell time includes. 
This picture is in line with Boltzmann's generalized Newtonian viscosity law
\cite{puscasu:jcp:2010,hansen:book:2022}, 
but different from the Maxwell-Weichert element. 
The inclusion of a correction function then leads to an extended single element generalized Maxwell model with 
only two free fitting parameters at a given characteristic wave-vector.  
The focus here is not on the dynamics of super-cooled liquids or glasses, see e.g. 
Refs. \cite{lemaitre:prl:2014,maier:jcp:2018} for more advanced theories, but on less viscous 
fluid systems where application of the Maxwell model is meaningful. 

The hydrodynamic function we study is the transverse velocity autocorrelation function (TVACF)
defined by 
$
	C_{uu}^\perp(k, t) = \frac{1}{V}\langle \widetilde{u}(k_y, t) \widetilde{u}(-k_y, 0)\rangle ,
$
where $V$ is the system volume, $\widetilde{u}$ is the spatial Fourier coefficient 
of the velocity $x$-component, and $k_y$ the $y$-component of the wave-vector 
$\vec{k}=(0, k_y, 0)$. $\langle \ldots \rangle$ denotes the ensemble 
average of uncorrelated initial conditions. Notice that to first order in the 
density fluctuations $C_{uu}^\perp$ is related to the transverse momentum autocorrelation function, 
$C_{jj}^\perp$, by $C_{jj}^\perp =  \rho^2 C_{uu}^\perp$, where $\rho$ is the average density. 
In the following we omit the subscript on the wave-vector $y$-component and set $k=k_y$. 

The hydrodynamic prediction for the TVACF is here derived, giving a foundation for the following 
discussion. We first write the momentum balance equation in Fourier space for $\vec{k}=(0, k, 0)$ 
and since we investigate equilibrium relaxations we can assume zero advection, thus,  
\begin{equation}
	\label{eq:P}
	\rho \partial_t \widetilde{u} = ik \widetilde{\sigma} \, .
\end{equation}
The Maxwell model introduces a characteristic elastic time and we only generalize with respect to wave-vector 
and not frequency. Moreover, we only consider homogeneous and isotropic systems. 
It is informative to write the generalized Maxwell model in real space
\begin{eqnarray}
	&\mbox{}&\eta_0 \int_{-\infty}^\infty f(y-y') \partial_{y'} u \, \d y'  = 
	\nonumber \\	
	&\mbox{}& \sigma(y,t) + \tau_M \partial_t \int_{-\infty}^\infty g(y-y')\sigma(y',t) \d y' \, ,
\end{eqnarray}
where $f$ and $g$ are the Maxwell viscosity and time kernels, respectively. 
The system response is thus given by two convolutions meaning that the
visco-elastic response is non-local. The kernels have dimensions of inverse length and must fulfill  
$
	\int _{-\infty}^\infty f(y-y') \d y' = \int _{-\infty}^\infty g(y-y') \d y' = 1.
$ 
Newton's law of viscosity is re-captured by ignoring both the elastic component, $\tau_M = 0$, 
as well as the spatial correlations, that is, $f(y-y')=\delta(y-y')$, where $\delta$ is the Dirac delta. 
Applying the convolution theorem the generalized Maxwell model reads in Fourier space
\begin{eqnarray}
	\label{eq:maxwellReciprocal}
	 i\eta_0k \widetilde{f}(k)\widetilde{u}(k) = 
	\left(1 + \tau_M \widetilde{g}(k) \partial_t\right) \widetilde{\sigma}(k,t)
\end{eqnarray}
with $\widetilde{f}(0)=\widetilde{g}(0)=1$. From this and 
the momentum balance equation, Eq. (\ref{eq:P}), we get 
\begin{equation}
	\partial^2_t\widetilde{u} + 
	\frac{1}{\widetilde{\tau}_M}\partial_t \widetilde{u} + 
	\frac{k^2\widetilde{\eta}_M}{\rho\widetilde{\tau}_M}\widetilde{u} = 0 \ ,
\end{equation}
where we have used the short hand notation $\widetilde{\eta}_M = \eta_0 \widetilde{f}(k)$ and 
$\widetilde{\tau}_M = \tau_M \widetilde{g}(k)$. Introducing the shear wave speed, $c_T(k)$, as
$
	c_T^2 = \widetilde{\eta}_M/\rho\widetilde{\tau}_M, \
$
multiplying by $\widetilde{u}(-k, 0)$, and ensemble averaging we get the dynamical 
equation the TVACF
\begin{equation}
	\label{eq:tvacfde}
	\partial_t^2 C_{uu}^\perp + 
	\frac{1}{\widetilde{\tau}_M}\partial _tC_{uu}^\perp + 
	c_T^2 k^2 C_{uu}^\perp = 0 \ , 
\end{equation}
for any wave-vector $k$. The eigenvalues (or characteristic frequencies) are
\begin{equation}
	\omega_{1,2} 
	= - 1/(2\widetilde{\tau}_M) \pm \sqrt{1/(2\widetilde{\tau}_M)^2 - (c_Tk)^2} \, . 
\end{equation}
In the visco-elastic regime $1/\widetilde{\tau}_M^2 < 4(c_Tk)^2$, or equivalently 
$1/\widetilde{\tau}_M < 4\widetilde{\eta}_M k^2/\rho$, the solution reads 
\begin{equation}
	\label{eq:tvacfSimple}
	C_{uu}^\perp(k,t)=\frac{k_BT}{\rho} e^{-\Gamma t} \cos \left(\omega_0 t\right) 
\end{equation}
by application of the equipartition theorem, and where we define the attenuation 
coefficient $\Gamma = 1/(2\widetilde{\tau}_M)$ and 
frequency $\omega_0^2 = |1/(2\widetilde{\tau}_M)^2 - (c_Tk)^2|$. It is clear that the frequency can be written 
as functions of $\widetilde{\tau}_M$,  or equivalently, via the attenuation coefficient
\begin{equation}
	\label{eq:frequency}
	\omega_0^2 = 4\Gamma|\Gamma - 2 \widetilde{\eta}_M k^2/\rho|. 
\end{equation}
Considering the original model it is natural to assume that $\widetilde{\eta}_M$ 
is simply the wave-vector dependent 
shear viscosity, $\widetilde{\eta}_0$. Indeed, we 
have $\lim_{k\rightarrow 0} \widetilde{\eta}_M = \lim_{k\rightarrow 0} \widetilde{\eta}_0 = \eta_0$. 
Therfore, if $\widetilde{\eta}_0$ is known, we obtain a single parameter model with $\Gamma$ (or equivalently 
$\widetilde{\tau}_M$) as fitting parameter. In this case
the attenuation determines the frequency and the Maxwell model exhibits what is
here called attenuation-frequency locking.

It is worth highligting that in the viscous regime $1/\widetilde{\tau}_M >
4\widetilde{\eta}_Mk^2/\rho$,  
the solution to Eq. (\ref{eq:tvacfde}) is topologically different  \cite{hirsch:book:2013} 
from Eq. (\ref{eq:tvacfSimple}). 
In particular,  we have the general solution 
$C_{uu}^\perp = C_1 e^{\omega_1 t} + C_2e^{\omega_2 t}$,
where $C_1$ and $C_2$ are integration constants, and $\omega_{1} \approx -1/\tau_M + \eta_0k^2/\rho$ and 
$\omega_2 \approx - \eta_0k^2/\rho$. Now, since $\lim_{k \rightarrow 0} C_{uu}^\perp = k_BT/\rho$ for all $t$ we must 
have that $C_1 = 0$ and $C_2 = k_BT/\rho$. This means that we re-capture the Newtonian model as $k \rightarrow 0$ 
\cite{hansen:book:2022}. In the following we investigate on the visco-elastic regime alone.

The prediction from the generalized Maxwell model can be compared to data from 
standard molecular dynamics simulations of a simple binary Lennard-Jones fluid; using the binary system 
the elastic part of the response can enhanced considerably compared to a single 
component system. Moreover, the system's A and B type particles have different
interaction parameters which avoids 
crystallization, see Refs. \cite{kob:prl:1994,schroder:jcp:2020} and supplementary information for details. 
The simulations are carried out in the canonical ensemble  
in the temperature range 0.45 - 1.0 $\epsilon/k_B$, where $k_B$ is Boltzmann's constant and $\epsilon$ 
the simulation energy scale. The mass density is the standard 1.2 $m/\sigma^3$, where $m$ and $\sigma$ are the
simulation mass and length scales. For the Lennard-Jones system
all quantities are given in terms of $\epsilon, m,$ and $\sigma$; 
as usual we omit writing these dimensions explicitly. The viscosity ranges from 15.3 at $T=1$ 
to approximately 3600 
at $T=0.45$. The TVACF is calculated directly from the microscopic definition of the streaming velocity, 
which to first order in density fluctuations reads 
$\widetilde{u}(k,t) = m/\rho \sum_i v_i(t)e^{-iky_i(t)}$, where $v_i$ is the particle velocity $x$-component and 
$y_i$ the particle $y$-position \cite{hansen:book:2022}. 
It is important to state that the TVACF has been compared to the results 
from the momentum current autocorrelation function, justifying the first order fluctuation approximation 
(at least away from the critical point in the thermodynamic phase space). 
Assuming $\widetilde{\eta}_M = \widetilde{\eta}_0$, the  
vector dependent viscosity can be calculated from the stress or the TVACF itself, 
see Ref. \cite{hansen:pre:2007}; we use here the the latter method 
and to highlight that it is the TVACF from simulation we use the subscript MD rather 
than $uu$, i.e., 
\begin{eqnarray}
	\label{eq:mdTVACF}
	\widetilde{\eta}_0(k) = \lim_{\omega \rightarrow 0} \frac{\rho}{k^2} \frac{C_{\text{MD}}^\perp(k, t=0) - 
	i\omega \widehat{C}_{\text{MD}}^\perp(k,\omega)}{\widehat{C}_{\text{MD}}^\perp(k,\omega)} \, ,
\end{eqnarray}
where $\widehat{C}_{\text{MD}}^\perp = \int_0^\infty C_{\text{MD}}^\perp(k, t) e^{-i\omega t} \, \d t$.

\begin{figure}[h]
	\begin{center}
		\bigskip		
		\includegraphics[scale=.3]{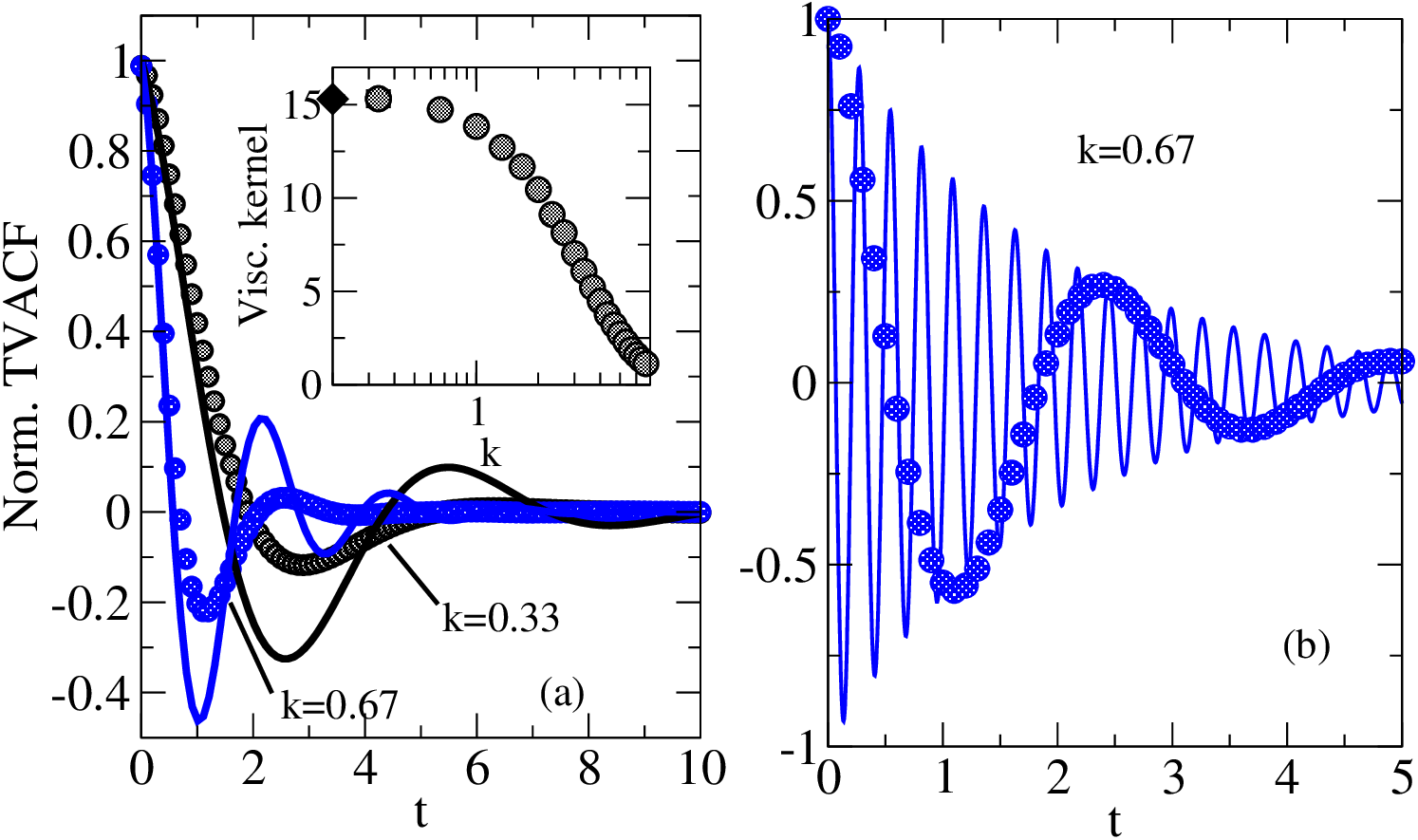} 
		\caption{\label{fig:CuuFitWithEta0Kernel}
			(a) The TVACF for the binary Lennard-Jones system at T=1.0. 
			Circles represent simulation data and full lines the least-squares 
			fit of Eq. (\ref{eq:tvacfSimple}) using $\Gamma$ 
			as fitting parameter. The inset shows the viscosity kernel - black diamond represent the 
			Green-Kubo shear viscosity value.	
			(b) Same as in (a), but for $T=0.45$ (viscosity kernel not shown for
			clarity, see supplemental material). 
		}
	\end{center}
\end{figure}

Figure \ref{fig:CuuFitWithEta0Kernel} (a) shows the least-squares fit of 
Eq. (\ref{eq:tvacfSimple}) to simulation data at $T=1.0$ using only $\Gamma$ as fitting parameter. 
The viscosity kernel, Eq. (\ref{eq:mdTVACF}), is used as independent input for
$\widetilde{\eta}_M$ and shown in the figure inset; 
for low wave-vector the kernel result is checked against the standard Green-Kubo integral of 
the stress auto-correlation function 
$\eta_0 = \frac{V}{k_BT}\int_0^\infty \langle \sigma(t)\sigma(0)\rangle \, \d t$. 
It is clear that due to the attenuation-frequency locking the wave-vector dependent Maxwell 
model fails to capture the correct relaxation dynamics when using the 
assumption $\widetilde{\eta}_M = \widetilde{\eta}_0$. The effect of the locking becomes 
even more extreme as we approach the super-cooled regime as shown in Fig.
\ref{fig:CuuFitWithEta0Kernel} (b); the attenuation is captured correctly, but
the frequency is not. This disagreement 
leads to the conjecture that $\widetilde{\eta}_M \neq \widetilde{\eta}_0$. 

Choosing the attenuation to be a free fitting parameter needs a comment. 
Since the viscosity kernel has reached the Green-Kubo limit at around $k=0.33$, one can expect the system 
to be in the wave-vector independent regime for $k \leq 0.33$. The fit in Fig. \ref{fig:CuuFitWithEta0Kernel} (a)
to the TVACF data yields a Maxwell relaxation time 
$\tau_M \approx \tau_M g(k=0.33) = 0.85$, whereas estimating this value from the stress auto-correlation function 
$\tau_M =  \eta_0/G$ with $G=V/k_BT\langle P_{xy}(0)P_{xy}(0)\rangle$, see Refs.  
\cite{heyes:prb:1988,hartkamp:pre:2013}, gives 0.24 \cite{noteontauM}. This further highlights 
the failure of the Maxwell model and why the Maxwell time is here estimated from 
fitting rather than from independent methods.

We now introduce a correction function, $w$, which is a measure of the 
deviation from the ideal mixing, specifically, it measures the system's deviation from the shear 
viscous response given by $\widetilde{\eta}_0$. In the general case, we must expect that 
the deviation is length scale dependent, 
i.e., $w$ is wave-vector dependent, and we have $\widetilde{\eta}_M=w(k)\widetilde{\eta}_0$. 
Notice that in real space $w$ is a kernel itself. The hydrodynamic equation for the TVACF is still given by 
Eq. (\ref{eq:tvacfSimple}), and the frequency can be written on the form, Eq. (\ref{eq:frequency}), 
$	\omega_0^2 = 4\Gamma \left| \Gamma - 2 w\widetilde{\eta}_0 k^2/\rho \right|$.
For a given wave-vector this is now a two-parameter model if the viscosity kernel is known.
Only the real valued correction is considered, and since we expect additional 
energy storage processes we expect the bounds $0 < w < 1$; 
in particular, as we approach the more elastic regime $w$ approaches 0.

\begin{figure}[h]
	\begin{center}
		\includegraphics[scale=.35]{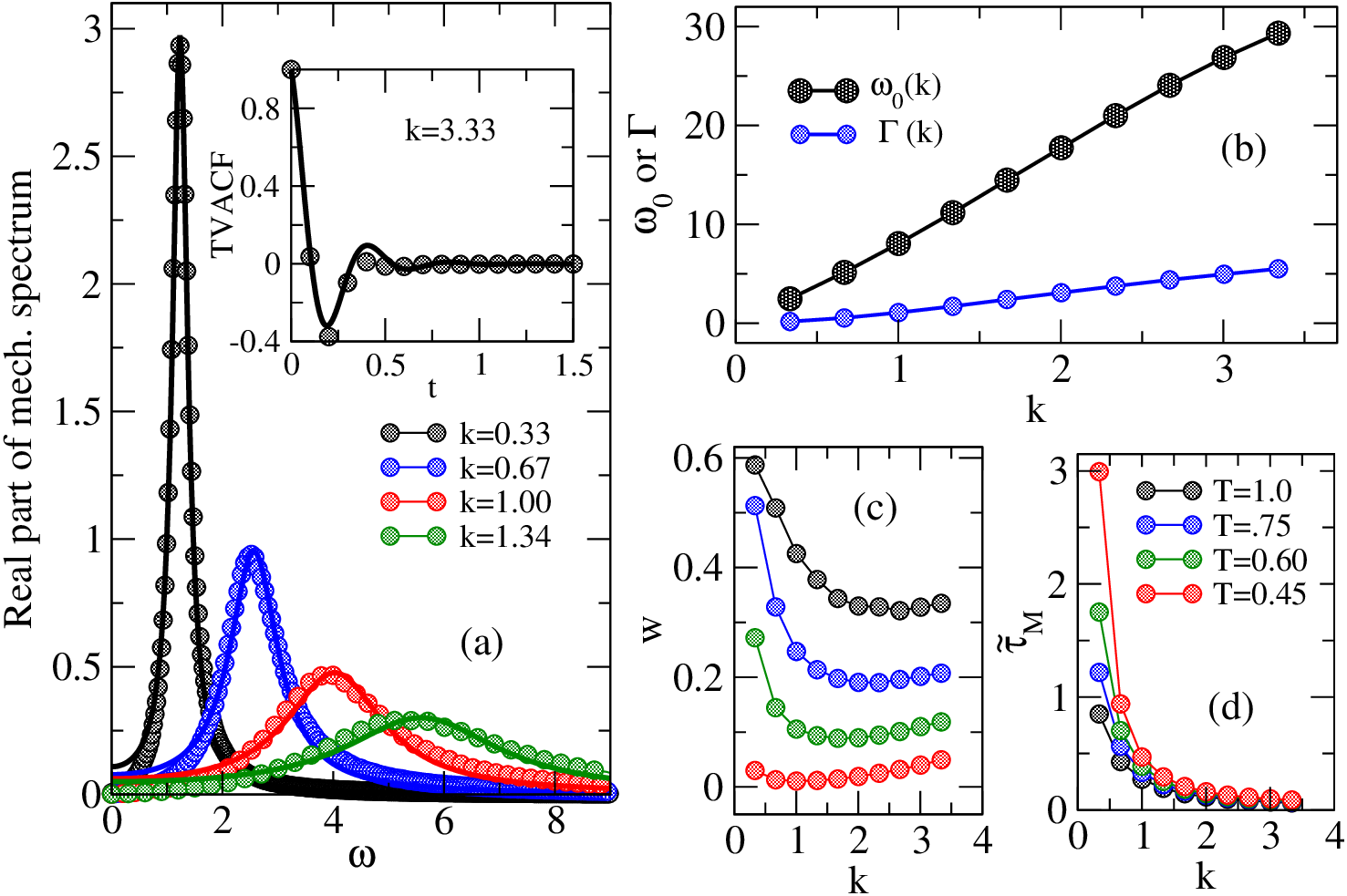}
		\caption{\label{fig:fitExtended}
			(a) Real part of the TVACF spectrum for the binary Lennard-Jones system 
			at $T=0.45$. Circles represent simulation data (Fourier-Laplace transformed) and full lines the 
			least squares fit of Eq. (\ref{eq:spectrum}). 
			Inset compares the corresponding agreement in time domain at highest
			wave-vector studied.
			(b) Dispersion plot for $\Gamma$ and $\omega_0$.
			(c) and (d) The correction function and Maxwell relaxation time as functions of wave-vector and 
			at different temperatures.
		}
	\end{center}
\end{figure}

Following the literature \cite{levesque:pra:1973,hansen:book:2006,mizuno:prl:2013}, 
we fit the corrected model to the real part of the corresponding TVACF spectrum; this enhances 
the deviation between theory and data. The spectrum is in terms of attenuation and frequency
\begin{eqnarray}
	\label{eq:spectrum}
	\widehat{C}_{uu}^\perp (k, \omega) = 
	\frac{k_B T}{\rho} \frac{i\omega + \Gamma}{(i\omega + \Gamma)^2 + \omega_0^2}
	\ .
\end{eqnarray}
Figure \ref{fig:fitExtended} (a) shows the least-squares fit of the real part of Eq. (\ref{eq:spectrum}) 
to simulation data for the binary Lennard-Jones system at $T=0.45$. The fitting parameters used are 
$\widetilde{\tau}_M$ and $w$. Again, $\widetilde{\eta}_0$ is found independently 
from Eq. (\ref{eq:mdTVACF}), see supplemental material.
As one expects, the agreement is not perfect, however, the corrected model does capture the 
main relaxation dynamics. The inset shows the TVACF for the largest wave-vector studied, namely $k=3.33$; 
this corresponds to a characteristic length scale of just 
1.89 or below two particle diameters. For larger wave-vectors the TVACF decays
extremely fast setting an upper bound for the spectral analysis. 
In Fig. \ref{fig:fitExtended} (b) the dispersion plot for 
the frequency and attenuation coefficients are shown. The linear 
dispersion for $\omega_0$ follows the results from Mizuno and Yamamoto \cite{mizuno:prl:2013}. 
The squared wave-vector dependency reported for $\Gamma$ is less clear with the wave-vectors studied 
here. This indicates that this two parameter model also captures 
the correct multi-scale dynamics. Figure \ref{fig:fitExtended} (c) shows that $w$ 
decreases as the temperature decreases. This is in agreement with the expectation above, 
namely, that the viscous response is reduced compared to the ideal mixing case.
Importantly, this reduction is present taking the wave-vector dependency into
account and in this sense it is an excess reduced viscous response. Moreover, the correction function 
features a minimum which shift towards lower $k$-values as the temperature decreases. For $T=0.45$ 
the minimum is located at approximately $k=1$, corresponding to a length scale of around
six particle diameters. This means that there exists a system characteristic length scale 
of maximum excess reduced viscous response that increases with decreasing
temperature. 
For completeness, the wave-vector dependent Maxwell time $\widetilde{\tau}_M$ is also
plotted, Fig. \ref{fig:fitExtended} (d). This shows the well-known "anomaly"
\cite{mizuno:prl:2013}: 
the Maxwell time is large on large length scales, and decreases abruptly as one approaches the 
microscopic length scale. 

\begin{figure}[t]
	\begin{center}
		\includegraphics[scale=0.325]{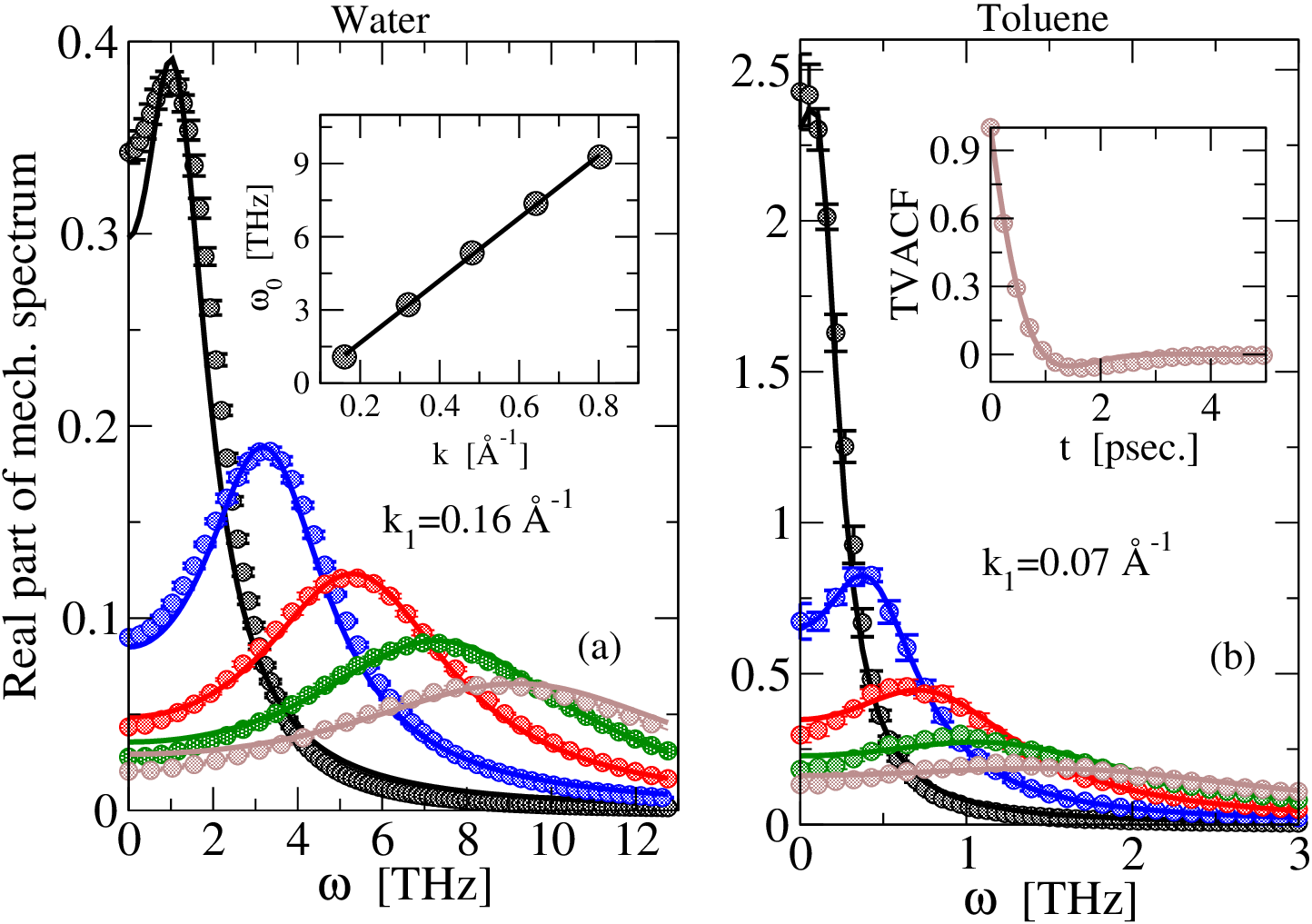}
		\caption{\label{fig:otherfluids}
			(a) Real part of the TVACF spectrum for water. Inset shows the dispersion 
			relation for $\omega_0$; line is represents a linear fit. 
			(b) As in (a), but for liquid toluene. Inset shows the TVACF (time
			domain) for $k=0.28$ {\AA}$^{-1}$. 
			In both (a) and (b) circles with lines represent data and full line fits 
			to Eq. (\ref{eq:spectrum}). 
			$k_1$ is the fundamental wave-vector (black) and the data/fits the subsequent modes.
		}
	\end{center}
\end{figure}

The corrected model is tested against simulation data for water, 
Fig. \ref{fig:otherfluids}(a), and liquid toluene, 
Fig. \ref{fig:otherfluids} (b), at ambient conditions. The water model is the flexible SPC/Fw model, 
see Ref. \cite{wu:jcp:2006}, and the toluene model is a united-atomic-unit model \cite{hansen:molsim:2021}. 
Again, see supplementary information for more details.
Clearly, the corrected model also performs well for these liquid systems, in particular, the linear dispersion 
relation for $\omega_0$ is present (shown for water). For water the viscous corrections 
varies from 0.54 to 0.34 in the wave-vector range $k$=0.16 - 0.8 {\AA}$^{-1}$ indicating that the viscous
response is significantly reduced for these more realistic model systems as well. 

In conclusion, the single element Maxwell model can successfully be extended using a correction function that, 
phenomenologically, includes the reduced viscous response not captured by the original model. 
With only two fitting parameters, namely, the Maxwell relaxation time and the correction 
function, the model predictions agree very well with data for the TVACF of different fluid systems. 
Moreover, the dispersion curve for the frequency will follow
previous results from a two Maxwell element (four parameters) model proposed by 
Mizuno and Yamamoto \cite{mizuno:prl:2013}. 
From the correction function it is concluded that the system viscous response is significantly
reduced. This reduction increases with decreasing temperature as expected, and, furthermore, 
the correction function features a minimum 
for low temperatures defining a length scale of maximum excess reduced viscous
response. The underlying microscopic processes behind this phenomenon and how it
relates to other theories is not clear at the present. 
It can also be concluded that the damped oscillations observed in the TVACF for non-viscous systems 
follows to a good approximation the functional form predicted by the single element Maxwell model, 
Eq. (\ref{eq:tvacfSimple}), and a better quantitative agreement is 
achieved by simply relaxing ideal mixing. Some complex and 
extreme viscous liquids feature mechanical spectra can differ qualitatively 
from the Maxwell model, see e.g. Ref. \cite{hecksher:jcp:2017}, 
and here more complex modelling effort must, of course, be undertaken.

The author thank Jeppe Dyre and Ulf R. Pedersen for fruitful discussions and
suggestions to this work.

\bibliographystyle{unsrt}

\begin{thebibliography}{10}

\bibitem{koplik:pfa:1989}
J.~Koplik, J.R. Banavar, and J.F. Willemsen.
\newblock Molecular dynamics of fluid flow at solid surfaces.
\newblock {\em Phys. Fluid. A}, 1:781--794, 1989.

\bibitem{travis:pre:1997}
K.~P. Travis, B.~D. Todd, and D.~J. Evans.
\newblock Depature from {N}avier-{S}tokes hydrodynamics in confined liquids.
\newblock {\em Phys. Rev. E}, 55:4288--4295, 1997.

\bibitem{hansen:book:2022}
J.~S. Hansen.
\newblock {\em Nanoscale Hydrodynamics of Simple Systems}.
\newblock Cambridge University Press, Cambridge, 2022.

\bibitem{thien:book:2002}
N.~Phan-Thien.
\newblock {\em {Understanding viscoelasticity: an introduction to rheology}}.
\newblock Springer Verlag, Berlin, 2002.

\bibitem{eijkel:mnf:2005}
J.C.T. Eijkel and A.~van~den Berg.
\newblock Nanofluidics: what is it and what can we expect from it?
\newblock {\em Microfluidics Nanofluidics}, 1:249--267, 2005.

\bibitem{bryk:jcp:2010}
T.~Bryk, I.~Mryglod, T.~Scopigno, G.~Ruocco, F.~Gorelli, and M.~Santoro.
\newblock Collective excitations in supercritical fluids: Analytical and
  molecular dynamics study of “positive” and “negative” dispersion.
\newblock {\em J. Chem. Phys.}, 133:024502, 2010.

\bibitem{bocquet:csr:2010}
L.~Bocquet and E.~Charlaix.
\newblock Nanofluidics, from bulk to interface.
\newblock {\em Chem. Soc. Rev.}, 39:1073, 2010.

\bibitem{bocquet:lapchip:2014}
L.~Bocquet and P.~Tabeling.
\newblock Physics and technological aspects of nanofluidics.
\newblock {\em Lab on a Chip}, 14:3143, 2014.

\bibitem{hansen:molsim:2021}
J.~S. Hansen.
\newblock {Where is the hydrodynamic limit?}
\newblock {\em Mol. Sim.}, 47:1391, 2021.

\bibitem{alley:pra:1983}
W.~E. Alley and B.~J. Alder.
\newblock Generalized transport coefficients for hard spheres.
\newblock {\em Phys. Rev. A}, 27:3158, 1983.

\bibitem{palmer:pre:1994}
P.J. Palmer.
\newblock Transverse-current autocorrelation-function calculations of the shear
  viscosity for molecular liquids.
\newblock {\em Phys. Rev. E}, 49:359, 1994.

\bibitem{levesque:pra:1973}
D.~Levesque, L.~Verlet, and J.~K\"{u}rkijarvi.
\newblock {Computer "Experiment" on Classical Fluids. IV. Transport Properties
  and Time-Correlation Functions of the Lennard-Jones Liquid near its Triple
  Point }.
\newblock {\em Phys.Rev.A}, 7:1690, 1973.

\bibitem{hansen:book:2006}
J.~P. Hansen and I.~R. McDonald.
\newblock {\em Theory of Simple Liquids}.
\newblock Academic Press, Amsterdam, 2006.

\bibitem{maxwell:1867}
J.C. Maxwell.
\newblock On the dynamical theory of gases.
\newblock {\em Phil. Trans. R. Soc. Lond.}, 157:49, 1867.

\bibitem{tschoegl:book:1989}
N.W. Tschoegl.
\newblock {\em {The Phenomenological Theory of Linear Viscoelastic Behavior}}.
\newblock Springer-Verlag, 1989.

\bibitem{mizuno:prl:2013}
H.~Mizuno and R.~Yamamoto.
\newblock {General Constitutive Model for Supercooled Liquids: Anomalous
  Transverse Wave Propagation}.
\newblock {\em Phys.Rev.Lett.}, 110:095901, 2013.

\bibitem{puscasu:jcp:2010}
R.M. Puscasu, B.D. Todd, Peter~J. Daivis, and J.S Hansen.
\newblock Nonlocal viscosity of polymer melts approaching their glassy state.
\newblock {\em J. Chem. Phys.}, 133:144907, 2010.

\bibitem{lemaitre:prl:2014}
A.~Lemaitre.
\newblock {Structural Relaxation is a Scale-Free Process}.
\newblock {\em Phys.Rev.Lett}, 113:245702, 2014.

\bibitem{maier:jcp:2018}
M.~Maier, A.~Zippelius, and M.~Fuchs.
\newblock Stress auto-correlation tensor in glass-forming isothermal fluids:
  From viscous to elastic response.
\newblock {\em J.Chem.Phys.}, 149:084502, 2018.

\bibitem{hirsch:book:2013}
M.W. Hirsch, S.~Smale, and R.L. Devaney.
\newblock {\em {Differential Equations, Dynamical Systems, and an Introduction
  to Chaos}}.
\newblock Elsevier, Oxford, 2013.

\bibitem{kob:prl:1994}
W.~Kob and H.C. Andersen.
\newblock Scaling behavoir in the beta-relaxation regime of a supercooled
  {L}ennard-{J}ones mixture.
\newblock {\em Phys.Rev.Lett.}, 73:1376, 1994.

\bibitem{schroder:jcp:2020}
T.B. Schr{\o}der and J.C. Dyre.
\newblock {Solid-like mean-square displacement in glass-forming liquids}.
\newblock {\em J. Chem. Phys.}, 152:141101, 2020.

\bibitem{hansen:pre:2007}
J.~S. Hansen, P.~J. Daivis, K.~P. Travis, and B.~D. Todd.
\newblock Parameterization of the nonlocal viscosity kernel for an atomic
  fluid.
\newblock {\em Phys. Rev. E}, 76:041121, 2007.

\bibitem{heyes:prb:1988}
D.~Heyes.
\newblock {Transport coefficients of Lennard-Jones fluids: A molecular-dynamics
  and effective-hard-sphere treatment}.
\newblock {\em Phys. Rev. B}, 37:5677, 1988.

\bibitem{hartkamp:pre:2013}
R.~Hartkamp, P.J. Daivis, and B.D. Todd.
\newblock Density dependence of the stress relaxation function of a simple
  liquid.
\newblock {\em Phys. Rev. E}, 87:032155, 2013.

\bibitem{noteontauM}
{At $T=1.0$ there does not exist a well-defined second plateau in the stress
  autocorrelation function, hence, the time zero phonon plateau is used. This
  follows Heyes and Hartkamp et al., see references}.

\bibitem{wu:jcp:2006}
Y.~Wu, H.~L. Tepper, and G.~A. Voth.
\newblock Flexible simple point-charge water model with improved liquid-state
  properties.
\newblock {\em J. Chem. Phys.}, 124:024503, 2006.

\bibitem{hecksher:jcp:2017}
T.~Hecksher, N.B. Olsen, and Jeppe~C. Dyre.
\newblock Model for the alpha and beta shear-mechanical properties of
  supercooled liquids and its comparison to squalane data.
\newblock {\em J.Chem.Phys.}, 146:154504, 2017.

\end{thebibliography}

\end{document}